\documentclass[12pt,preprint]{aastex}

\usepackage{rotate,epsfig,natbib}
\newcommand{\spit}{\textit{Spitzer Space Telescope}}

\newcommand{\one}{3.6~micron}
\newcommand{\two}{4.5~micron}
\newcommand{\three}{5.8~micron}
\newcommand{\four}{8.0~micron}

\newcommand{\rsxot}{0.97}
\newcommand{\erspxot}{+0.02}
\newcommand{\ersmxot}{-0.02}

\newcommand{\rpxot}{0.98}
\newcommand{\erppxot}{+0.03}
\newcommand{\erpmxot}{-0.03}

\newcommand{\inclxot}{88.9}
\newcommand{\einclpxot}{+0.7}
\newcommand{\einclmxot}{-0.7}

\newcommand{\semixot}{0.0369}
\newcommand{\esemixot}{0.0002}

\newcommand{\flone}{0.00081}
\newcommand{\eflone}{0.00017}
\newcommand{\hjdone}{2454421.10118}
\newcommand{\ehjdone}{0.01675}
\newcommand{\dtone}{-6.9}
\newcommand{\edtone}{24.1}

\newcommand{\fltwo}{0.00098}
\newcommand{\efltwo}{0.00020}
\newcommand{\hjdtwo}{2454423.72707}
\newcommand{\ehjdtwo}{0.01306}
\newcommand{\dttwo}{7.8}
\newcommand{\edttwo}{18.8}

\newcommand{\flthr}{0.00167}
\newcommand{\eflthr}{0.00036}
\newcommand{\hjdthr}{2454421.10719}
\newcommand{\ehjdthr}{0.01315}
\newcommand{\dtthr}{1.8}
\newcommand{\edtthr}{18.9}

\newcommand{\flfour}{0.00133}
\newcommand{\eflfour}{0.00049}
\newcommand{\hjdfour}{2454423.71817}
\newcommand{\ehjdfour}{0.01221}
\newcommand{\dtfour}{-5.1}
\newcommand{\edtfour}{17.6}

\newcommand{\twosouth}{0.00064} 
\newcommand{\etwosouth}{0.00033}
\newcommand{\timesouth}{+11.1}
\newcommand{\etimesouth}{24.8}
\newcommand{\twoearth}{2.4}

\newcommand{\foursig}{-1.4}

\newcommand{\twosign}{ 1.6}
\newcommand{\threesign}{ 2.2}

\newcommand{\ecoswaug}{0.012}

\newcommand{\fernper}{2.6158640}
\newcommand{\ferneper}{0.0000016}
\newcommand{\fernhjd}{2,454,466.88467}

\newcommand{\gradone}{-0.011}
\newcommand{\egradone}{0.005}
\newcommand{\gradtwo}{-0.010}
\newcommand{\egradtwo}{0.004}

\newcommand{\timeper}{6.0}

\renewenvironment{thebibliography}[1]{%
\begin{oldthebibliography}{#1}%
\setlength{\parskip}{0ex}%
\setlength{\itemsep}{0ex}%
}%
{%
\end{oldthebibliography}%
}

\slugcomment{Accepted for publication in The Astrophysical Journal}

\shorttitle{\objectname[NAME XO-2]{XO-2}Thermal emission}
\shortauthors{Machalek et al.}

\begin{document}


\title{Detection of Thermal Emission of XO-2b: Evidence for a Weak Temperature Inversion}


\author{
Pavel~Machalek\altaffilmark{1,2},
Peter~R.~McCullough\altaffilmark{2},
Adam~Burrows\altaffilmark{3},
Christopher~J.~Burke\altaffilmark{4},
Joseph~L.~Hora\altaffilmark{4},
Christopher~M.~Johns-Krull\altaffilmark{5}
}

\email{pavel@jhu.edu}
\altaffiltext{1}{Department of Physics and Astronomy, Johns Hopkins University, 3400 North Charles St., Baltimore MD 21218}
\altaffiltext{2}{Space Telescope Science Institute, 3700 San Martin Dr., Baltimore MD 21218}
\altaffiltext{3}{Department of Astrophysical Sciences, Princeton University, Princeton, NJ 08544}

\altaffiltext{4}{Harvard-Smithsonian Center for Astrophysics, 60 Garden St., Cambridge, MA 02138} 
\altaffiltext{5}{Department of Physics and Astronomy, Rice University, 6100 Main Street, MS-108, Houston, TX 77005}

\begin{abstract}
We estimate flux ratios of the extrasolar planet XO-2b to its host star \objectname[NAME XO-2]{XO-2}~at 3.6, 4.5, 5.8 and 8.0 micron with IRAC on the \spit~to be \flone~$\pm$~\eflone, \fltwo~$\pm$~ \efltwo, \flthr~$\pm$~\eflthr~and ~\flfour~$\pm$~\eflfour, respectively.  
The fluxes provide tentative evidence for a weak temperature inversion in the upper atmosphere, the precise nature of which would need to be confirmed by longer wavelength observations.
\objectname[NAME XO-2b]{XO-2b}~substellar flux of 0.76$\times$ 10$^{9}$ ergs cm$^{-2}$ s$^{-1}$ lies in the predicted transition region between atmospheres with and without upper atmospheric temperature inversion.

\end{abstract}


\keywords{stars:individual(XO-2) --- binaries:eclipsing --- infrared:stars --- planetary systems}



\section{Introduction}
The field of comparative planetology has burgeoned in the past year \citep{deming_review09}: four exo-planet secondary eclipse observations have been published with all four IRAC channels: HD 209458b~\citep{knutson07b}; HD 189733b~\citep{charb08};  XO-1b~\citep{machalek08} and TrES-4~\citep{knutson_tres4}, which were  used to deduce the temperatures structure in the planetary upper atmospheres. 
Numerous scientific firsts were also announced: a model dependent derivation of the radiative time constant of an extrasolar planet atmosphere \citep{laughlin09}, first unambiguous detection of water vapor in a Hot Jupiter atmosphere  in HD 189733b \citep{grill09}, together with prospects for detailed study of extrasolar planet atmospheres with JWST \citep{seager08}. 
\\

Planetary stratospheres, which are common in the giant planets of the Solar System \citep{depater01}, form when high stellar flux penetrates deep into the atmosphere where the atmosphere does not efficiently radiate the heat away. Hot-Jupiters were also observed to possess stratospheres \citep{knutson07b,machalek08,knutson_tres4} and enhanced opacity at high altitude in the form of an extra absorber of optical light has been suggested as the physical cause of upper atmospheric temperature inversions (\citet{hubeny03,burr07b,fort07b}). 
\citet{spiegel09} have recently argued against vanadium oxide (VO) as the extra optical absorber and suggested that even the previously favored gaseous optical absorber titanium oxide (TiO) would probably rain out of the upper atmosphere unless unusually high levels of macroscopic mixing exist (Eddy diffusion coefficients K$_{zz}$ $\sim$ 10$^{7}$-10$^{11}$ depending on particulate size of the TiO condensates) compared to Jupiter ( K$_{zz}$ $\sim$ 10$^{6}$ \citep{depater01}) to overcome gravitational settling. 
\citet{zahn09} has considered the role by which absorption of UV and visible light by S$_{2}$ and S$_{3}$ can lead to formation of thermal inversion in upper atmospheres of Hot-Jupiters and suggested planetary metallicity, in addition to substellar flux,  as the determining factor for the presence of a temperature inversion.  \\

\begin{figure}[!th]
  \centering
  \includegraphics[width=0.6\textwidth]{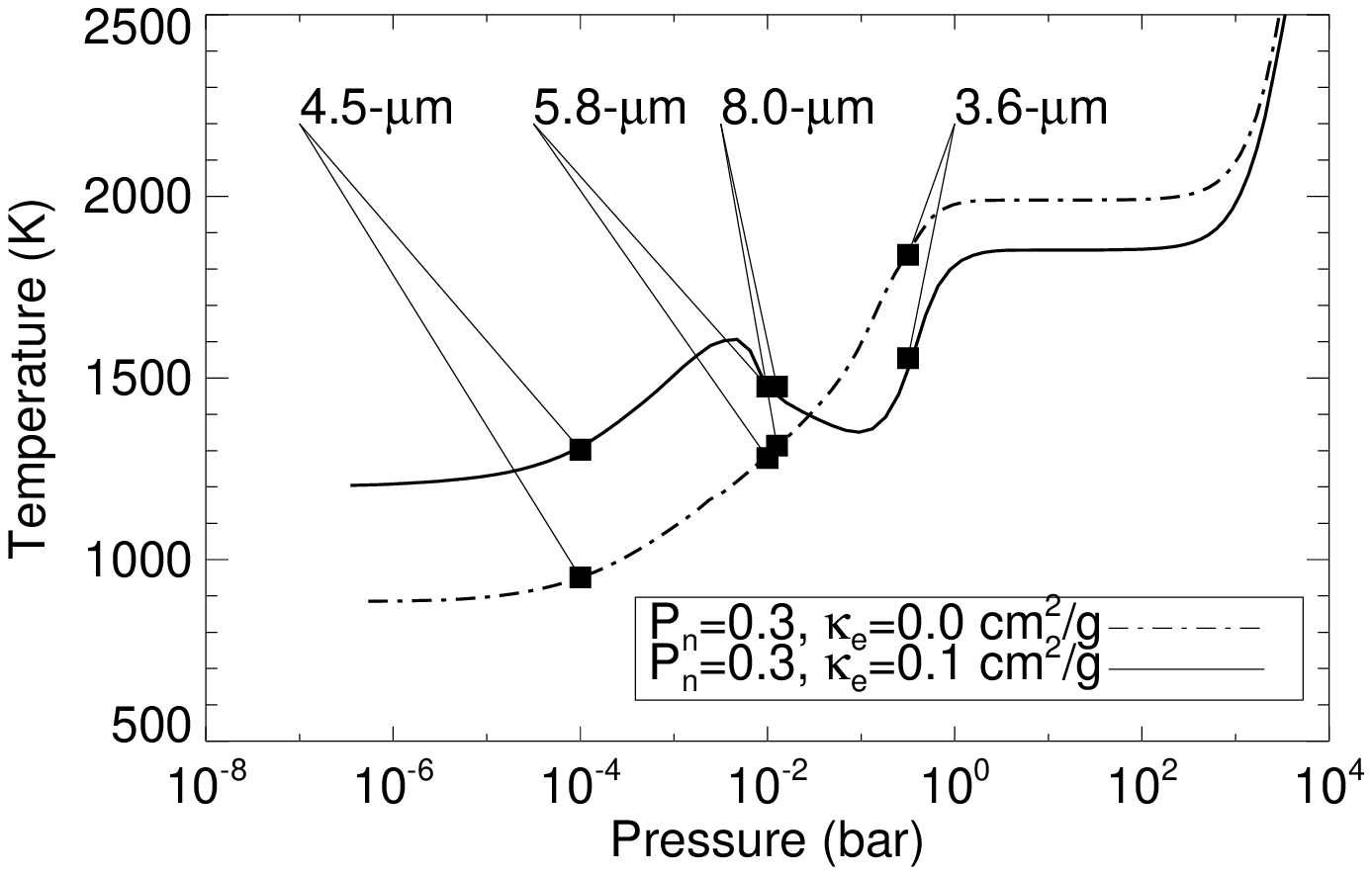}
  \caption{(top) Temperature / Pressure profiles for the atmosphere of XO-2b following the methodology of \citet{burr07,burr07b,spiegel09} for heat redistribution parameter P$_{n}$=0.3 with no upper atmospheric optical absorber (dot-dashed line) and a corresponding model with a uniform upper atmospheric absorber (solid line) (absorption coefficient $\kappa_{e}$=0.1 cm$^{2}$/g) with depths corresponding to emission in the IRAC channels denoted. \label{xo2b_tp} }
\end{figure}

Hot-Jupiter atmosphere models \citep{burr07,burr07b,spiegel09,fort07b} show that water and CO opacity define the $\tau$=2/3 decoupling layers and the relative temperatures at those layers determine the relative brightness at \one~and \two. In the presence of an optical absorber in the upper atmosphere the \two~planet flux is higher than the \one~flux and vice-versa for models without an optical absorber in the upper atmosphere, which serves as an important diagnostic for detection of thermal inversions in upper atmospheres of Hot-Jupiters (see Fig. \ref{xo2b_tp} for the XO-2b Temperature / Pressure profile).
 

A detailed study of the IR secondary eclipse planetary spectra of \objectname{HD 209458b}, \objectname{HD 189733b}, \objectname[NAME TrES-1b]{TrES-1}, \objectname{HD 149026b} and non-eclipsing \objectname{HD 179949b}, and \objectname[* ups And b]{$\upsilon$ And b} by \citet{burr07b} suggests that the presence of such an upper atmospheric absorber might be dependent on the flux from the star at the sub-stellar point on the planet as well as second order effects like metallicity and planetary surface gravity. 
In the \citet{burr07b} interpretation planets with high sub-stellar point flux (e.g., \objectname{HD 209458b}, OGLE-Tr-56b, OGLE-Tr-132b, \objectname[GSC 03549-02811]{TrES-2b} and \objectname[GSC 03727-01064]{XO-3b}) would have extra optical absorber in the upper atmospheric layer and water features in emission while planets with lower fluxes (\objectname[NAME XO-1b]{XO-1b},~\objectname[NAME TrES-1b]{TrES-1},~\objectname[TYC 3413-5-1]{XO-2b} and \objectname{HD 189733b}) would have no such extra absorber and would possess water features in absorption. \citet{fort07b} also suggest a similar division of planetary spectra based on incident stellar flux.
Observations of XO-2b and other planetary systems directly constrain and test the incident flux threshold necessary for the occurrence for a thermal inversion in the upper atmosphere of a Hot-Jupiter. 

Based on the planetary sub-stellar point flux from the star, both \citet{burr07b} and \citet{fort07b} predicted that \objectname[NAME XO-1b]{XO-1b}~ should not have exhibited a thermal inversion in its upper atmosphere, yet a temperature inversion is observed \citep{machalek08}. 
The incident flux on \objectname[NAME XO-2b]{XO-2b} from its parent star (0.76$\times$ 10$^{9}$ ergs cm$^{-2}$ s$^{-1}$) lies in the predicted transition region between upper atmospheres with or without thermal inversions as delineated by \citet{burr07b,fort07b} . 
\citet{burr07b} predicted a ``weak stratosphere'' for XO-2b.\\


\objectname[NAME XO-2]{XO-2}~(\objectname{2MASS J07480647+5013328}) has high metallicity, [Fe/H]=0.45~$\pm$~0.02, high proper motion, $\mu_{tot}$=157 mas yr$^{-1}$, and a common proper motion stellar companion with 31" separation \citep{burke07}.  
The planet ~\objectname[TYC 3413-5-1]{XO-2b} has an orbital period of~\fernper~$\pm$~\ferneper~days \citep{fern09} and orbits around the northern declination component of the  ~\objectname[TYC 3413-5-1]{XO-2} stellar binary system (see Fig. \ref{fig:gs}) . We present observations of the infrared spectral energy distribution (SED) of the planet ~\objectname[TYC 3413-5-1]{XO-2b}~  \citep{burke07} in all 4 IRAC channels obtained during secondary eclipses observed with  \spit. By comparing XO-2b's SED with atmospheric models, we test for the presence of a thermal inversion in the upper atmosphere of \objectname[NAME XO-2b]{XO-2b}.

\section{Observations}
\begin{figure}[!h]
\centering
\includegraphics[width=0.5\textwidth]{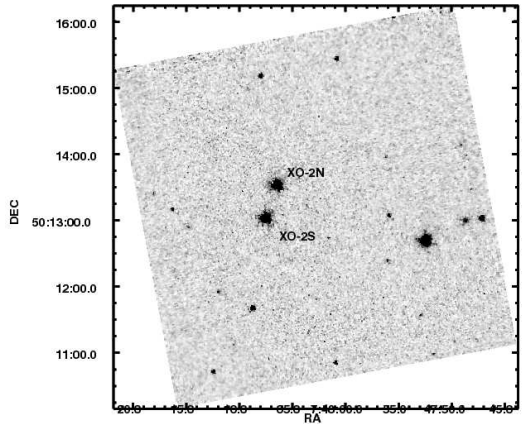}
\caption{The field-of-view of the XO-2 binary system in the IRAC 3.6 micron channel. North is up, East is to the left. XO-2N hosts the transiting extrasolar planet XO-2b. \label{fig:gs} }
\end{figure}

The InfraRed Array Camera \citep[IRAC;][]{fazio04} has a field of view of
5.2$\arcmin$ $\times$ 5.2$\arcmin$ in each of its four bands. Two adjacent
fields are imaged in pairs (3.6 and 5.8 microns; 4.5 and 8.0 microns). The
detector arrays each measure 256 $\times$ 256 pixels, with a pixel size of
approximately 1.22$\arcsec$ $\times$ 1.22$\arcsec$. 
We have observed the 
~\objectname[TYC 3413-5-1]{XO-2b}~ system in all 4 channels in two separate Astronomical
Observing Requests (AORs) in two different sessions: the 3.6 and 5.8 micron
channels for 5.94 hours on UT 2007 November 16 (AOR 24462080) and the 4.5 and
8.0 micron channels for 5.94 hours on UT 2007 November 19 (AOR 24462336). We used
the full array 2s+2s/12s frame time in the stellar mode in which the \one~and \two~bands are exposed for two consecutive 2s exposures while the \three~and \four~bands are integrating for 12s to prevent detector saturation.
Figure \ref{fig:gs} shows a representative IRAC \one~image.  

We used the standard IRAC Basic Calibrated Data (BCD) products (version 16.1)
described in the Spitzer Data
Handbook\footnote{\url{http://ssc.spitzer.caltech.edu/irac/dh/}},
which includes dark frame subtraction, multiplexer bleed correction,
detector linearization, and flat-fielding of the images.  The starting point for our analysis were the BCD images. We converted the times recorded by the spacecraft in the FITS file header keyword DATE-OBS
to heliocentric Julian dates using the orbital ephemeris of the spacecraft
provided by the Horizons Ephemeris System\footnote{\url{http://ssd.jpl.nasa.gov/}}. We flagged cosmic ray pixels in the images, which resulted in [0.8; 0.4; 2.4; 3.1] \% of photometric points in the [3.6; 4.5; 5.8 and 8.0] micron time series to have a flagged pixel in the photometric aperture and which were not used in the analysis. We did not resample the pixels in any way during our analysis doing so could compromise the photometry. 

To determine the scalar background for the photometry we subtracted the zodiacal background in each channel by constructing a histogram of all pixels in each image and fitting a Gaussian to the distribution of the zodiacal background brightness. 
We evaluated the centroids of both North and South components of ~\objectname[TYC 3413-5-1]{XO-2b}~ with the IDL CNTRD routine which locates where the second derivatives of x and y pixel position reach 0. We have found that CNTRD centroids have lower rms in both x and y pixel directions than the centroids produced by IDL routine GCNTRD, which fits Gaussians to marginal x and y distributions.  The pointing  in the \one~channel varied by $\sim$~0.10 pixels in x and $\sim$~0.06 pixels peak-to-peak in y and by $\sim$~0.15 pixel in x and $\sim$~0.20 pixel in y in the \two~channel. The shifting of the
stellar centroid within a pixel, which have sub-pixel sensitivity
variations, resulted in a modulation of the stellar flux in the \one~and \two~channels (described below).\\



\subsection{3.6 and 4.5 micron time series}
\label{insb}

\begin{figure}[!h]
\centering
\includegraphics[width=0.5\textwidth]{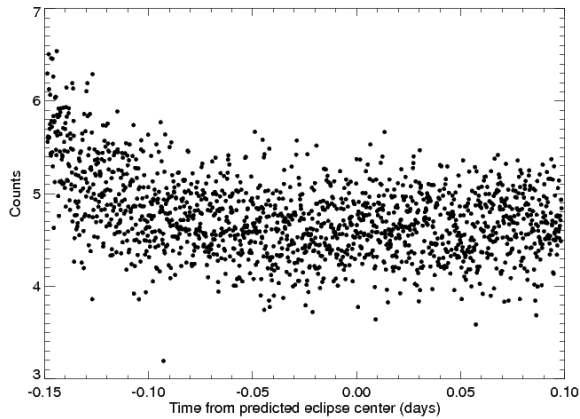}
\caption{The average sky count rates in the annulus with radii between 13 and 50 pixels around the star XO-2N for the 3.6 micron IRAC channel observations. Notice the asymptotic increase of the sky count rates in the first $\sim$ 1 hour of observation. Similar trend was observed for the other binary component XO-2S.}
\label{fig:sky}
\end{figure}
We performed aperture photometry on the \one~and \two~background subtracted images with aperture radii ranging from 2.5 to 5.0 pixels in 0.5 pixel increments for both components of the XO-2 binary. 
The sky level was elevated during the first hour of observations in the \one~channel (see Fig. \ref{fig:sky}). This behavior is opposite the asymptotic behavior in the \one~channel for the similarly bright star TrES-4~\citep{knutson_tres4}.  The \one~ and \two~time series exhibited a sharp increase during the first $\sim$ 20 minutes of exposure for \objectname[NAME XO-2]{XO-2}~and the 2 calibrators, presumably  as a result of the instrument reaching a new equilibrium after previous observations. Such relaxation effects can reach several percent and usually stabilize within the first hour of observations of a new target. We have rejected the first 180  points  in the \one~and \two~time series.

 The aperture radius was selected by choosing the lowest rms for out-of-eclipse points.  Aperture of 2.5 pixels was chosen for the \one~time series, with an out-of-eclipse rms of 0.0044 after systematic effect removal which is 1.02 times the photon noise limit based on stellar brightness, background flux and detector read noise. 
Similarly, for the \two~series an aperture of 2.5 pixels resulted in the lowest out-of-eclipse rms of 0.0063 which 1.05 times the photon noise limit.
We applied the appropriate aperture correction for each channel to the stellar flux value according to the Spitzer Data Handbook of [1.112, 1.113, 1.125, 1.218] for the [3.6, 4.5, 5.8 and 8.0] micron channels, respectively. 

\begin{figure}
\centering
\includegraphics[totalheight=0.9\textwidth]{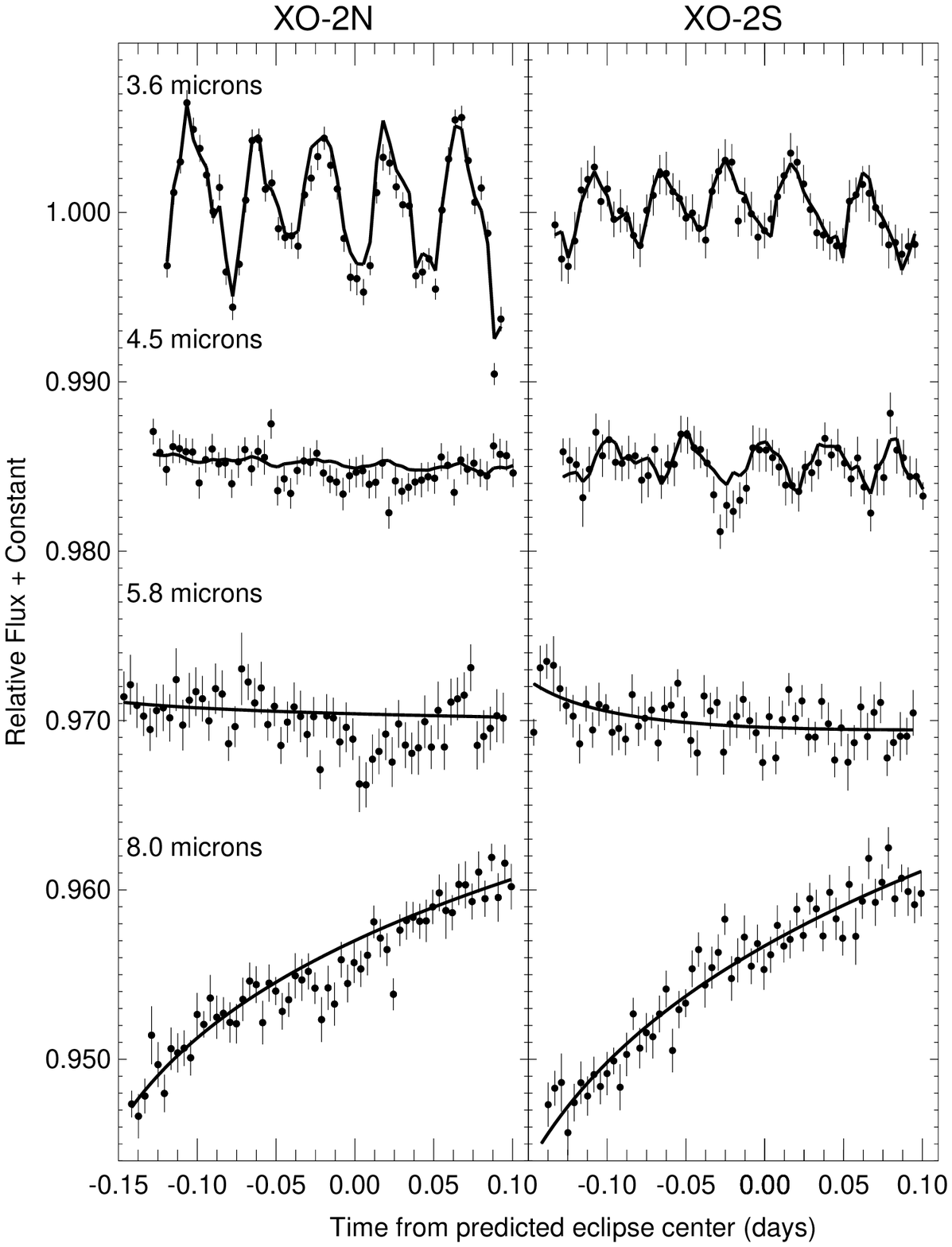}
\caption{\small (left) Secondary eclipse observations of \protect\objectname{XO-2b}~ with 
IRAC on \spit~ in \one, \two, 
\three~and \four~channels (from top to bottom) binned in \timeper-minute intervals 
and normalized to 1 and offset for clarity. XO-2b orbits XO-2N the northern component of XO-2 binary. (right) The light curve of the southern component XO-2S of the binary system (with no planet) plotted to the same scale and included for comparison. The overplotted solid  lines do \textit{not} represent a fit to the time series, but rather show the 
corrections for the detector effects (see text).}
\label{fig:instru}
\end{figure}

A strong correlation between the sub-pixel centroid and stellar brightness was observed in both the \one~and \two~channels, with flux amplitudes of $\sim$ 1.3 \% and 0.5 \%, respectively. This well studied effect \citep{charb05,morales06,machalek08,knutson_tres4} is due to the InSb detector intrapixel sensitivity variations as the spacecraft jitters $\sim$0.1 - 0.3 arcsec in orientation over a period of $\sim$3000 seconds\footnote{\url{http://ssc.spitzer.caltech.edu/documents/exoplanetmemo.txt}}. The uncorrected sub-pixel intensity variations are clearly visible in the time series of \objectname[NAME XO-2]{XO-2}~in the \one~and \two~channels in Fig.~\ref{fig:instru}.\\

Despite the similarity between the XO-2 stellar binary components (same brightness J = 9.74$\pm$~0.02 and similar color (J-H)$_{NORTH}$ = 0.40$\pm$~0.03 vs. (J-H)$_{SOUTH}$ = 0.37$\pm$ 0.03) the amplitude of their the sub-pixel sensitivity variations in the \one~channel for the North and South  component of ~\objectname[TYC 3413-5-1]{XO-2b}~ is markedly different (1.3 \% for XO-2N vs 0.7\% for XO-2S, see Fig.~\ref{fig:instru}). This is most likely due to the different positions of the two XO-2 components with respect to the edge of the pixel and hence being affected  differently by the pixel response function. 
\citet{laughlin09} have noted in their 30-hour long \four~time series of the HD80606 and HD80607 binary components, which also have same brightness and similar color, that due to different sub-pixel positions the detector systematic ramps are different. 

 We have corrected for the sub-pixel intensity variations by fitting a function of 4 variables to the time series of \objectname[NAME XO-2b]{XO-2b}~: 

\begin{equation}
  \label{eq:subpixel}I_{subpixel}=~ b_{1} +~b_{2}x +~b_{3}y~+~b_{4}t, 
\end{equation}
 where b$_{i}$ are coefficients; $x$ and $y$  are subpixel centroids of the stellar flux and t is the time from the start of observations in days. We have tried  quadratic and cross terms for the subpixel centroids correction \citep{desert09} but that has not improved our $\chi ^{2}$ and hence were not used.    
We fit the secondary eclipse light curves using the formalism of \citet{agol02} with no stellar limb darkening as appropriate for a secondary eclipse. We adopt stellar and planetary parameters from \citet{burke07} \footnote{We have redone the entire analysis using the updated planetary and stellar parameters of \citet{fern09} and our results do not change withing uncertainties.}
: R$_{\star}$ = \rsxot $^{\erspxot}_{\ersmxot}$ R$_{\sun}$, R$_{p}$ = \rpxot $^{\erppxot}_{\erpmxot}$ R$_{Jup}$\footnote{1 R$_{Jup}$ = 71,492 km.}, $i$ =~\inclxot$^{\einclpxot}_{\einclmxot}$ degrees, and $a$ = \semixot~$\pm$~\esemixot~AU with updated ephemeris from \citet{fern09}:

\begin{equation}
  \label{eq:ephm} T_{c}(E) = \fernhjd0 (HJD) + E(\fernper~days) \, .
\end{equation}

\begin{figure}
\centering
\includegraphics[totalheight=0.9\textwidth]{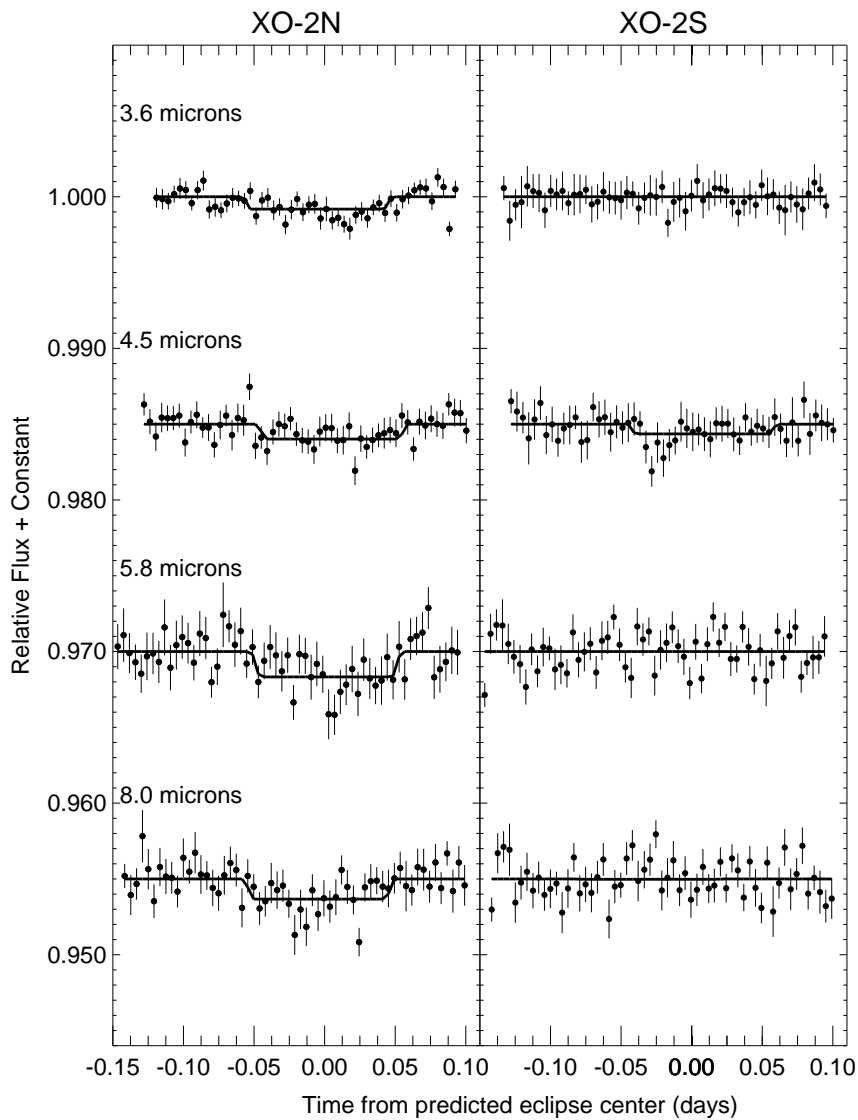}

\caption{(left) Secondary eclipse of \protect\objectname{XO-2b} around the binary component XO-2N observed with IRAC on \spit~in 3.6, 4.5, 5.8, and 8.0 micron channels (top to bottom) corrected for detector effects, normalized and binned in \timeper-minute intervals and offset for clarity. The best-fit eclipse curves are overplotted. (right) The time series of the southern component XO-2S of the XO-2 binary, which has no eclipsing planet, after detector correction. The non-zero depth in the \two~channel of XO-2S  is a spurious residual of systematic effects that mimics an eclipse. See \S \ref{insb} for details.}
\label{fig:fit}
\end{figure}

We have fitted the 4 baseline detrending parameters (a constant, linear x-position, linear y-position and linear time terms) concurrently with the depth of the eclipse $\Delta F$ and the timing of the centroid $\Delta T$  (total of 6 fitting parameters) using  a Monte Carlo Markov Chain (MCMC) with 10$^{5}$ iterations such that the ratio of jumps for each parameter was between 20-40 \% (see \citet{gregory05,mark09}). 
The initial 20\%  iterations were rejected to remove arbitrary starting conditions and the final parameters were chosen as the median value from the posterior probability distribution for each parameter. 
The adopted eclipse depth $\Delta F$ and centroid timing $\Delta T$ in minutes from the expected secondary eclipse mid-center time for an assumed eccentricity of zero are reported in Table \ref{tbl1} with errors obtained from the symmetric 66.8 \% contours around the median of the posterior probability distribution from the MCMC runs. 
Best-fit eclipse curves binned in \timeper-minute intervals are plotted in Fig.~\ref{fig:fit} and the eclipse parameters are listed in Table \ref{tbl1}. They are the channel wavelength, eclipse depth $\Delta F$, eclipse mid-center time in HJD and the timing offset  $\Delta T$.

We find that the \one~time series of ~\objectname[TYC 3413-5-1]{XO-2b}~ exhibits a linear flux trend with a slope of b$_{4}$ = \gradone \% $\pm$ \egradone \% per hour. This trend is consistently removed from our time series by including a linear time term b$_{4}$ as one of the MCMC 6 parameter fit, as described above.   \citet{knutson_tres4} noted a linear trend in their \one~time series of TrES-4 with a slope of +0.030  $\pm$ 0.004 \%~per hour and attributed such  linear trend to a previously unknown instrumental effect of the detector. The \two~time series of ~\objectname[TYC 3413-5-1]{XO-2b}~ exhibits a similar linear trend to the \one~time series with a slope of \gradtwo \% $\pm$ \egradtwo \% per hour, which is lower than a corresponding slope for TrES-4 of -0.020$\pm$0.003\%~per hour \citep{knutson_tres4} even though XO-2N is 11 times brighter. 


To assess the amount of non-Gaussian correlated red noise left in the time series after systematic effect removal we have performed the complete MCMC analysis on the other binary component XO-2S in the XO-2 binary system, which does not have a short-period Jovian mass orbiting it (which would have easily been detected by radial velocity observations) and thus does not exhibit a secondary eclipse. 
The best fit eclipse curves of XO-2S are plotted on the right in Fig.~\ref{fig:fit}. The \one, \three~and \four~channels exhibit eclipse depths consistent with zero  while the \two~channel shows an eclipse with depth \twosouth~$\pm$~\etwosouth~based on Gaussian statistics with a time shift of \timesouth~$\pm$~\etimesouth~min. 
We exclude the possibility that the spurious eclipse at \two~around XO-2S could be due to a transiting super-Earth (corresponding to R$_{p}$=\twoearth R$_{\earth}$ around XO-2S) by noting that the eclipse is not observed in the \four~time series of XO-2S, which allows for maximum depths of 0.00002 $\pm$ 0.00001 and since it was observed at the same time as the \two~time series we can reject the transiting super-Earth hypothesis at more than 10-$\sigma$.  

The significant non-zero eclipse depth around the control star XO-2S which does not have short-period Jovian-mass planet suggests that the \two~time series for both XO-2N and XO-2S underestimates the amount of time correlated red-noise even after our calibration procedures. 

\subsection{5.8 and 8.0 micron time series}
\label{sias}
Aperture photometry was performed on the images with cosmic ray pixels flagged  with an aperture radius between 2.5 and 6.0 pixels in 0.5 pixel increments. The optimal size of the aperture was determined by minimizing the rms scatter in the light curve for observations outside of the eclipse. Aperture of 3.5 pixels was chosen for the \three~time series, with an out-of-eclipse rms of 0.0066 which is 32 \% greater than the the theoretical  noise  based on stellar brightness, background flux and detector read noise. 
Similarly for the 8.0 micron series an aperture of 4.0 pixels resulted in the lowest out-of-eclipse rms of 0.0054 which is 35\% higher than the photon noise limit.  
The internal scattering of photons inside \three~and \four~Si:As arrays is likely  responsible for the fact that we do not approach the Poisson limit as closely as in the 3.6~micron and 4.5~micron channels.\footnote{\url{http://ssc.spitzer.caltech.edu/documents/irac\_memo.txt}} 
Unlike for the XO-1b observations \citet{machalek08}, where the first $\sim$ 30 minutes of observations were rejected as the instrument settled into a new equilibrium state, we do not observe such an initial ramp-up in the \three~and \four~time series and conversely we do not reject any data points from the beginning of the time series in these channels. 

 Fig.~\ref{fig:instru} shows intensity variation with time, which is caused by changes in the effective gain of individual pixels over time. This effect has been observed before by \citet{dem05}, both in the IRAC camera and in the IRS and MIPS 24~micron cameras and is dependent on the illumination level of the individual pixel \citep{knutson07,knutson07b}. 
Pixels with high illumination will reach their equilibrium within $\sim$1 hour, but lower illumination pixels increase in intensity over time, approximately proportional to the inverse of the logarithm of illumination.  
 The detector ramp intensity in the \three~time series decreased in flux during the $\sim$ 6 hours of observation by $\sim$  0.1 \% for XO-2N and by $\sim$~0.5 \% for XO-2S, following a similar trend seen by \citet{knutson07b} in their \three~time series of brighter \objectname{HD 209458b}. 

 We detect a nonlinear increase in the brightness of \objectname[TYC 3413-5-1]{XO-2N}~and XO-2S in the \four~channel similarly to  \citet{knutson07} who have reported a nonlinear flux increase over time with the \four~IRAC detector.
Most recently the gain variation with time and illumination has been confirmed in the extended duration (30 hours) observations of HD80606 and HD80607 \citep{laughlin09}, who found that even for similarly bright and similar color binary, the \three~and \four~time series ``ramp'' can be different, presumably due to different position of the stars in relation to the edge of the pixels.  

To properly account for the way in which the non-linear flux ramps affect our estimate of the secondary eclipse depth and timing we have performed MCMC simultaneous fitting of the 3 ramp parameters (see Eq. \ref{eq:ramp}) and the eclipse depth $\Delta F$ and centroid timing $\Delta T$ (5 parameter fit overall):

\begin{equation}
  \label{eq:ramp}I_{model} = a_{1}  + a_{2} \times~  ln(\Delta t +0.05) + a_{3} \times~ (ln(\Delta t +0.05))^{2},
\end{equation}
where $I_{model}$ is the normalized model flux, $\Delta t$ is the time in days since the beginning of observations (the 0.05 factor is included to prevent singularity at $\Delta t$ =0 )  and $a_{i}$ are coefficients.

The MCMC fitting was performed in the same way as for the \one~and \two~channels with uncertainties obtained from the symmetric 66.8 \% contours around the median of the posterior probability distribution from the MCMC runs. 
The eclipse depth in the \three~channel of \flthr~$\pm$~\eflthr~represents an overall 4.7-$\sigma$ detection of the secondary eclipse based on the cumulative SNR of all the in-eclipse points and similarly the \four~channel eclipse of~\flfour~$\pm$~\eflfour~has an overall SNR of 2.7. 
The resultant time series were normalized using out-of-eclipse points and binned into \timeper-minute bins (Fig.~\ref{fig:fit}) for viewing clarity.


\section{Discussion}
\label{anl}

 To check whether our results depend on the aperture radius we have redone the MCMC analysis for photometry with aperture radii between 2.5 and 4.5 pixels for the \one~ and \two~time series and aperture radii from 3.0 to 5.0 pixels in the \three~and \four~time series and obtained consistent results for the eclipse depth $\Delta F$ and centroid timing $\Delta T$ within errors to our adopted values from Table \ref{tbl1}.

To test the robustness of our data reduction and MCMC analysis technique
and consistency with other observations in the IRAC full-array mode we
have reanalyzed the IRAC secondary eclipse
time series of \objectname[NAME XO-1]{XO-1} from \citet{machalek08} in all 4 IRAC channels with the new MCMC pipeline. The eclipse depths $\Delta F$  of XO-1b with our MCMC pipeline are still in agreement  with a model of a thermal inversion and an extra upper atmospheric absorber of uniform opacity of $\kappa _{e}$ = 0.1 cm$^{2}$/g and redistribution parameters of P$_{n}$ = [0.3] at [1.8,0.8,0.4]-$\sigma$ at \one,\two~and \three~bands, respectively. 
The secondary eclipse of XO-1b in the current MCMC analysis of \four~time series is 3.8-$\sigma$ above the thermal inversion model above but clearly inconsistent at 16.1-$\sigma$ with an atmospheric model without a temperature inversion. 
Our conclusion from \citet{machalek08} for a temperature inversion in the upper atmosphere of XO-1b caused by an optical absorber of uniform opacity of $\kappa _{e}$ = 0.1 cm$^{2}$/g is thus reinforced.  
 We have thus demonstrated that our MCMC reduction and analysis pipeline is robust and the secondary eclipse depth estimates are consistent with previous pipeline versions from \citet{machalek08}.  \\

 The eclipse mid-center timings for \objectname[NAME XO-2b]{XO-2b}~ in Table \ref{tbl1} are individually  consistent within uncertainties with zero timing residuals for a circular orbit based on the ephemeris by \citet{fern09} and a combined mid-eclipse timing offset of -1.0 $\pm$ 9.7 min. Using the equation of (e.g. \citet{kopal59} Eq. 9.23):
  
\begin{equation}
  \label{eq:ecc} e \times cos(\omega) \simeq \frac{\pi \Delta t}{2P}, 
\end{equation}
where $e$ is the eccentricity, $\omega$ is the longitude of periastron, $P$ is the orbital period, and $\Delta t$ is the centroid time shift from expected time of secondary eclipse, allows us to set a 3-$\sigma$ upper limit on e$\times$cos($\omega$)~$<$~\ecoswaug.

\begin{figure}
\centering
\includegraphics[angle=-270,totalheight=8cm]{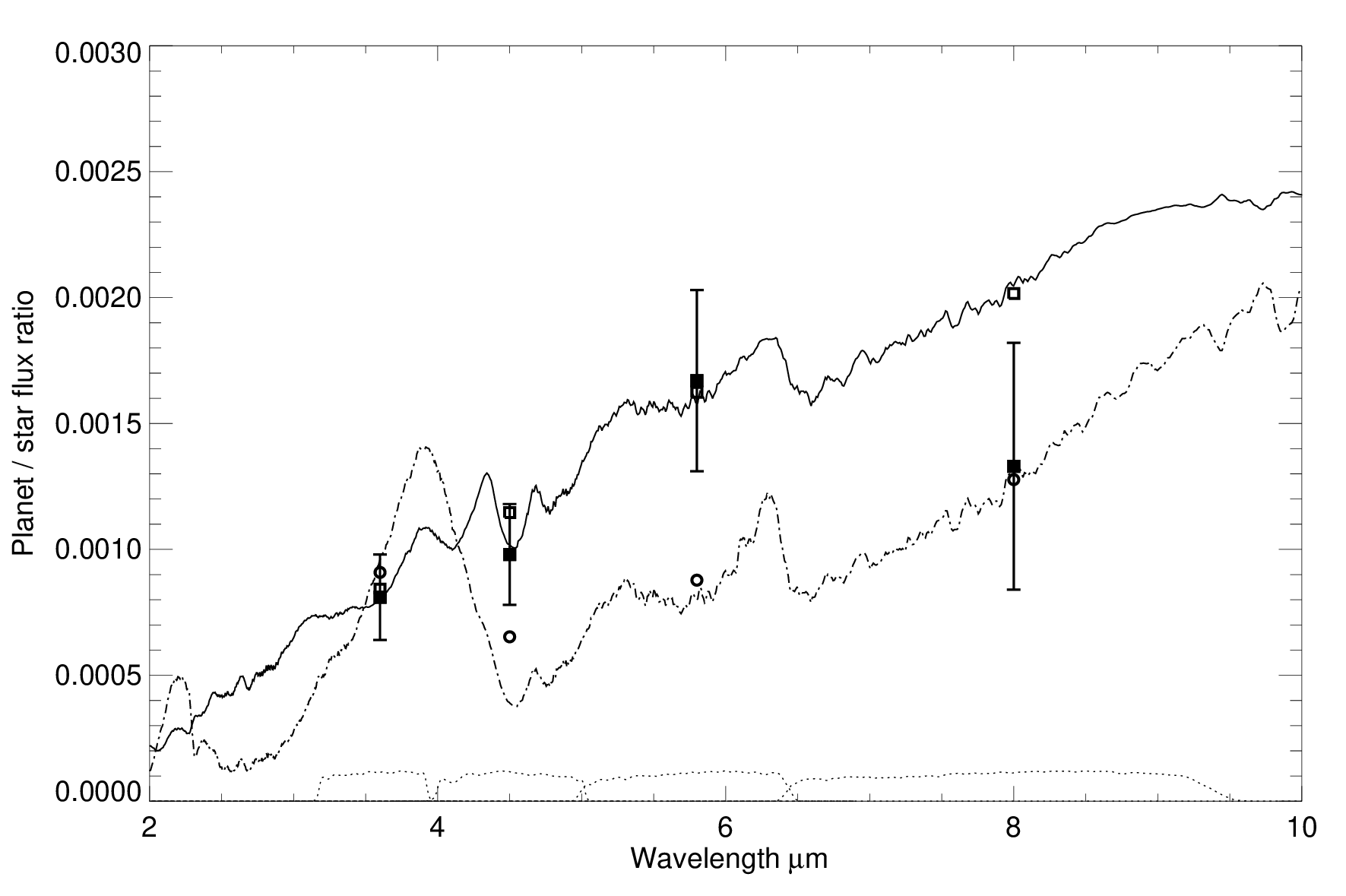}
\caption{~\spit~IRAC secondary eclipse depths for \protect\objectname{XO-2b} with MCMC error bars (filled squares). The predicted emission spectrum of the planet \citep{burr07,burr07b,spiegel09} with an upper atmospheric absorber of $\kappa_{e}$ = 0.1 cm$^{2}$/g and redistribution parameter of P$_{n}$=[0.3] is plotted as a solid line. 
A model with no atmospheric absorber and a redistribution parameters of P$_{n}$=[0.3] is over plotted with dot-dashed line (see \S \ref{anl} for details). The band-averaged flux ratios are plotted as open squares and open circles for the models with and without an extra upper atmospheric absorber, respectively. The normalized~\spit~IRAC response curves  for the 3.6-, 4.5-, 5.8-, and 8.0~micron channels are plotted at the bottom of the figure (dotted lines).
}
\label{fig:atmo}
\end{figure}

The \objectname[NAME XO-2b]{XO-2b}~ eclipse depths in Fig.~\ref{fig:atmo} show several trends.
The flux ratio of \objectname[NAME XO-2b]{XO-2b}~to the star peaks in the \three~channel, with a decrease towards the \one~ and \two~channels and a slight decrease towards
the \four~channel. 
%
The solid line and band averages represented as open squares in Fig.~\ref{fig:atmo} depict an atmospheric model of \objectname[NAME XO-2b]{XO-2b}, following the methodology of \citet{burr07,burr07b,spiegel09} with a thermal inversion and an extra upper atmospheric absorber of uniform opacity of $\kappa _{e}$ = 0.1 cm$^{2}$/g and redistribution parameters of P$_{n}$ = [0.3]. P$_{n}$=0 corresponds to no heat redistribution from the planetary day-side to the night-side and P$_{n}$ = 0.5 stands for full redistribution (see \citet{burr07b} for details).
The upper atmospheric temperature inversion model fits the observed eclipse depths within 1.0-$\sigma$ in the \one,~\two~and~\three~bands, respectively, while the \four~channel eclipse depth is offset by \foursig-$\sigma$ from the model. 
The model without an upper atmospheric temperature inversion with the redistribution parameter P$_{n}$ = [0.3] (dot-dashed curves) does not fit the observations by [\twosign,\threesign]-$\sigma$, respectively, in the \two~and \three~channels while fitting the \one~and \four~band eclipse depth within 1.0-$\sigma$. 
The eclipse depth fit in the \one, \two~and \three~bands thus provides evidence for a  ``weak'' temperature inversion in the upper atmosphere of XO-2b. XO-2b would thus be the second Hot-Jupiter along with XO-1b with a sub-stellar point flux of less than  $\sim$1.0 $\times$ 10$^9$ erg cm $^{-2}$ s $^{-1}$ to possess a temperature inversion in its upper atmosphere. 

The possibility of thermal inversion in a planetary upper atmosphere has been suggested by \citet{hubeny03,iro05,burr06,burr07} and \citet{fort07b}. 
A thermal inversion in the planetary upper atmosphere has been invoked for interpretation of broadband IR spectra in the case of \objectname{HD209458b} \citep{knutson07b}; XO-1b \citep{machalek08}; \objectname{HD149026b} \citep{har07} and recently TrES-4 \citep{knutson_tres4}. The presence of an extra optical absorber in the upper atmosphere, of yet unknown composition, would yield a thermal inversion in the planetary upper atmosphere and the presence of the water features in emission. 

 \citet{burr07b} and \citet{fort07b} suggested that the presence of the extra optical absorber in the upper atmosphere might be correlated with the incident flux from the star at the sub-stellar point on the planet. \objectname[NAME XO-2b]{XO-2b}~substellar flux of (0.76$\times$ 10$^{9}$ ergs cm$^{-2}$ s$^{-1}$) lies in the transition region between atmospheres with or without upper atmospheric temperature inversions. XO-1b which has an even lower substellar flux of 0.49$\times$ 10$^{9}$ ergs cm$^{-2}$ s$^{-1}$ yet possesses unambiguous signs of a temperature inversion in the upper atmosphere, which suggests that secondary effects like metallicity \citep{zahn09} and planetary surface gravity may also determine the  nature of the temperature inversion of the upper atmospheres of Hot-Jupiters.\\
 
Confirmation of the weak temperature inversion of XO-2b would be possible with longer wavelength observations because the contrast between models with or without a thermal inversion is high (see Fig.~\ref{fig:atmo}).
Further study of Hot-Jupiter upper atmospheres, especially of planets that lie in the predicted transition region of the substellar flux ($\sim$ 0.6 - 1.0$\times$ 10$^{9}$ ergs cm$^{-2}$ s$^{-1}$) like HAT-P-1, WASP-2b, HD197286,  should refine the sub-stellar flux boundary with respect to the presence/absence of an upper atmospheric thermal inversion. \\

\section{Conclusion}
We report the estimated flux ratios of the planet \objectname[NAME XO-2b]{XO-2b}~ in the \spit~IRAC 3.6, 4.5, 5.8 and 8.0 micron channels. 
The fluxes are consistent with a weak temperature inversion in the upper atmosphere of XO-2b.
The atmospheric model with an upper atmospheric temperature inversion with an extra optical absorber of opacity of $\kappa _{e}$ = 0.1 cm$^{2}$/g and redistribution parameter $P_{n}$=0.3 provides the best fit to the data in the \one, \two~and \three~channels with a mild \foursig-$\sigma$ inconsistency in the \four~channel.   


The presence or absence of the stratospheric absorber and thermal inversion layer has been linked to the flux from the parent star at the sub-stellar point on the planet. The \objectname[NAME XO-2b]{XO-2b}~sub-stellar point flux of $\sim$0.76 $\times$ 10$^9$ erg cm $^{-2}$ s $^{-1}$ is the second lowest so far reported for a planetary atmosphere with a thermal inversion. 

\acknowledgments
The authors would like to thank H. Knutson and N. Iro for helpful discussions. The authors would also like to acknowledge the use of publicly available routines by Eric Agol and Levenberg-Marquardt least-squares minimization routine MPFITFUN  by Craig Markwardt.  P.M. and P.R.M. were supported by the Spitzer Science Center Grant C4030 to the Space Telescope Science Institute.
A.B. was supported in part by
NASA grant NNX07AG80G.  We also acknowledge support through
JPL/Spitzer Agreements 1328092, 1348668, and 1312647. 
 This work is based on observations made with the Spitzer Space Telescope, which is operated by the Jet Propulsion Laboratory, California Institute of Technology under a contract with NASA. This publication also makes use of data products from the Two Micron All Sky Survey, which is a joint project of the University of Massachusetts and the Infrared Processing and Analysis Center/California Institute of Technology, funded by the National Aeronautics and Space Administration and the National Science Foundation.\\



\begin{deluxetable}{ccccc}
\tablecolumns{5}
\tablewidth{0pt}
\tablecaption{Secondary eclipse best fit parameters }
\tablehead{
\colhead{$\lambda$}  &\colhead{Eclipse Depth $\Delta F$} &\colhead{Eclipse Center Time} &\colhead{Time offset $\Delta T$ } \\
\colhead{(microns)}  &\colhead{} &\colhead{(HJD)} &\colhead{(min)}}
\startdata
3.6  &\flone~$\pm$ \eflone & \hjdone~$\pm$ \ehjdone& \dtone~$\pm$ \edtone\\
4.5  & \fltwo~$\pm$ \efltwo & \hjdtwo~$\pm$ \ehjdtwo& \dttwo~$\pm$ \edttwo\\
5.8  & \flthr~$\pm$ \eflthr & \hjdthr~$\pm$ \ehjdthr& \dtthr~$\pm$ \edtthr\\
8.0  & \flfour~$\pm$ \eflfour & \hjdfour~$\pm$ \ehjdfour& \dtfour~$\pm$ \edtfour\\
\enddata
\label{tbl1}
\end{deluxetable}

\clearpage

\clearpage




\end{document}